# Hyperbolic Metamaterial Nanoparticles for Efficient Hyperthermia in the II and III Near-Infrared Windows

*Yingqi Zhao[1], Marzia Iarossi[1,2], Nicolò Maccaferri[1,3], Lieselot Deleye[1], Giovanni Melle[1,2], Jian-An Huang[1], Francesco Tantussi[1], Tommi Isoniemi[1] and Francesco De Angelis[1]\**

[1] Istituto Italiano di Tecnologia, Via Morego 30, 16163, Genova, Italy

[2] Department of Informatics, Bioengineering, Robotics and Systems Engineering (DIBRIS), Università degli Studi di Genova, Via Balbi 5, 16126 Genova, Italy

[3] Physics and Materials Science Research Unit, University of Luxembourg, L-1511 Luxembourg, Luxembourg

## Abstract

The use of gold nanoparticles for hyperthermia therapy in near infrared (NIR) spectral regions has catalysed substantial research efforts due to the potential impact in clinical therapy applications. However, the photoscattering effect scaling with the square of the nanoparticle volume leads to a low absorption efficiency, which has hindered the utility of gold nanoparticles in NIR II regions above 1000 nm. Here, we conquer this limit by introducing hyperbolic metamaterial nanoparticles that are made of multi-layered gold/dielectric nanodisks and exhibit >70% absorption efficiency in the NIR II and III regions. Their high light-to-heat conversion is demonstrated by a much larger temperature increase than that of gold nanodisks with the same amount of gold. Efficient *in vitro* hyperthermia of living cells with negligible cytotoxicity shows the potential of our approach for next-generation bio-medical applications.

## Introduction

Over the last decade, photoreactive nanoparticles for bio-medical applications[1] have received significant attention due to their unique optical properties and their potential use in next generation theragnostic biotechnologies, such as photothermal therapy[2-3], photoacoustic imaging[4-5] and drug delivery[6]. Photothermal therapy, which exploits localized heat converted from near infrared (NIR) light, has been demonstrated to be a very promising low-invasive treatment for cancer cell ablation and theranostic applications.[7-12] Optical transparency windows for biological applications, such as NIR I and NIR II (located in the spectral ranges of 650-950 nm and 1000-1350 nm, respectively), provide optimal tissue treatment conditions.[13] In particular, the NIR II window offers more efficient penetration and less photon scattering in biological tissues; therefore, it is more preferred for photothermal therapy and photoacoustic imaging.[14-16] More recently, a third transparency window (NIR III) at 1600-1870 nm is emerging due to its deeper penetration depth and lower photon scattering in various types of tissues, including the brain.[17-18]

In this framework, intensive research efforts have been made to design novel materials and architectures that can be used as transducers to provide highly efficient light to heat conversion processes. Toward this aim, it is necessary to maximize absorbance in the NIR windows while limiting the scattering effects, which decrease the light-to-heat conversion efficiency.[19-20] Many examples of materials responsive in the NIR spectral region can be found, including copper sulfide-based materials[12], carbon nanotubes,[21] gold nanoparticles[22], silver nanoplates,[23] and polymer-based particles.[24] Among them, plasmonic gold nanoparticles represent a primary choice due to their biocompatibility and enhanced absorbance at their localized surface plasmon resonance (LSPR) wavelength in the NIR I window.[2, 25-26] Their LSPR can be tuned by modifying their sizes and/or geometry to match the illumination wavelength for an optimized hyperthermia performance.[27]

However, gold nanoparticles face intrinsic limitations in the NIR II and III regions when they tune the LSPR to a long wavelength by increasing their sizes.[22, 28] For example, for nanospheres that are smaller than the illumination wavelength, their absorption scales with their volume ($V$), whereas their the scattering scales with $V^2$.[29-30] Consequently, the scattering effect is dominant over the absorption for large nanoparticles, leading to low absorption efficiencies in the NIR II and III regions. Therefore, their application in the NIR II window is limited to few examples, including Au nanoshells,[13] nano nanoechinus,[31] and Au blackbody.[22] Nanoparticles operated at the NIR III window are reported even less. Alternative structure designs and fabrication approaches to conquer this limit are desirable in order to have a high absorption efficiency in the NIR II and III bands.

In this paper, we demonstrate that hyperbolic metamaterial (HMM) nanoparticles are high-efficient transducers in photothermal treatments in the NIR II and III windows. Their hyperbolic nature[32] enables a high confinement of the electric field within the nanostructures, which leads to a high absorption efficiency.[33] As a result, the light-to-heat conversion efficiency is increased with respect to conventional nanostructures made of the same amount of gold, thus making this class of metamaterials a promising candidate for photothermal applications. For simplicity, we will refer to these materials as hyperbolic meta-particles (HMPs). *In vitro* experiments are also conducted to demonstrate the superior hyperthermia performance of HMPs, which do not show cytotoxic effects, over gold nanodisks. Moreover, the pure absorption of the HMPs in the NIR III region suggests its potential in next generation light-to-heat transducers, photothermal therapy and photoacoustic imaging.

## Materials and Methods

### Simulations

To evaluate the optical and thermal responses of our meta-particles, we performed steady-state simulations with COMSOL®. For the electromagnetic simulations we used the RF Module.

The refractive index values of gold were taken from literature.[34] The refractive index values for the oxide were set to n = 1.5 and k = 0.7. We considered a simulation region where we specified the background electric field (a linearly polarized plane wave with an irradiance of 0.1 mW/μm²), and then calculated the scattered field by a single meta-particle to extract the optical parameters, such as the absorption and scattering cross sections. The model computes the scattering, absorption and extinction cross-sections of the particle on the substrate. The scattering cross-section is defined as $\sigma_{scat} = \frac{1}{I_0} \iint (\boldsymbol{n} \cdot \boldsymbol{S}) \, dS$, where $I_0$ is the intensity of the incident light, $\boldsymbol{n}$ is the normal vector pointing outwards from the nanodot and $\boldsymbol{S}$ is the Poynting vector. The integral is taken over the closed surface of the meta-antenna. The absorption cross section equals $\sigma_{abs} = \frac{1}{I_0} \iiint Q \, dV$, where $Q$ is the power loss density of the system and the integral is taken over the volume of the meta-antenna. The simulated steady-state temperature distribution reached under irradiation of 0.1 mW/μm² was retrieved using the COMSOL Multiphysics "Heat Transfer in Solids" module. For the glass substrate, a thermal conductivity of 1 Wm⁻¹K⁻¹ was considered, and the surrounding medium was assumed to be air. The baseline temperature of all simulations was ambient temperature.

**HMM nanoparticle fabrication**

HMPs were prepared via inductively coupled plasma (ICP) etching of the gold/silica multilayers using the Cr disk as a mask fabricated by hole mask colloidal lithography.[35] The flow chart of the fabrication process in shown in Figure S1. The microscope glass slides were first cleaned by sonicating in acetone and 2-propanol for two minutes each, and then the slides were washed with deionized water (DI water) and blow dried under an $N_2$ flow. Then, the glass slides were loaded into an electron beam deposition chamber (Ebeam, PVD75 Kurt J. Lesker company) to deposit the stacking Au/SiO₂ metal dielectric bi-layer consisting of 2 nm Ti + 10 nm Au and 2 nm Ti + 20 nm SiO₂. The deposition of the bi-layer units were repeated four times to achieve fabrication of the multilayers. On the top of the multilayers, photoresist (950 PMMA A8, Micro Chem) was spin coated at 6000 rpm and soft baked at 180°C for 1 min. After $O_2$ plasma treatment (2 min, 100 W, Plasma cleaner, Gambetti), Poly(diallyldimethylammonium chloride) solution (PDDA, Mw 200,000-350,000, 20 wt. % in $H_2O$, Sigma, three times diluted) was drop coated and incubated for 5 min to create a positively charged polymer layer. Then, the extra PDDA solution was washed away under flowing DI water. Then, negatively charged polystyrene (PS) beads (diameter 552 nm, 5 wt% water suspension, Micro Particle GmbH) were drop-coated on the as-prepared multilayers, washed under flowing DI water and dried with an $N_2$ flow. Therefore, randomly distributed PS beads were attached on the top of photoresist. The density of beads can be controlled by the deposition time and beads concentration. The beads size can be reduced by $O_2$ plasma etching in the inductively coupled plasma-reactive ion etching system (ICP-RIE, SENTECH SI500). Gold film (40 nm) was sputter coated (Sputter coater, Quorum, Q150T ES) on top of the sample to serve as an etching mask to protect the photoresist underneath. After removal of the PS beads by a polydimethylsiloxane (PDMS) film, the samples were treated again by $O_2$ plasma in an ICP-RIE system to remove the photoresist. The diameter of the holes was controlled by adjusting the size of the PS beads by a variable $O_2$ plasma treatment time. With the hole mask, 100 nm of chromium (Cr) was deposited in the Ebeam system with a vertical incident angle. After photoresist lift-off in acetone, randomly distributed Cr disks on the stacking multilayer were fabricated as a final etching mask of the multilayers. ICP-RIE etching was conducted with a $CF_4$ gas flow of 15 sccm, a radio frequency (RF) power of 200 W, an ICP power of 400 W, a temperature of 5°C, and a pressure of 1 Pa. The etching time was adjusted according to the stacking film thickness to ensure all the extra Au and $SiO_2$ was removed. Then, the Cr was removed by soaking the sample in Cr etchant (Etch 18, Organo Spezial Chemie GmbH) for 2 min. After cleaning with DI water and drying under an $N_2$ flow, the sample morphology was characterized by scanning electron microscopy (SEM).

To fabricate a control sample of the gold nanodisks, hole masks with the same hole density and diameter as the HMPs were fabricated directly on the glass substrate. Through the mask, 40 nm of Au were deposited with 2 nm Ti as an adhesion layer at the bottom. After lift-off in acetone, the gold nanodisk array supported by a glass substrate with the same amount of the gold as the HMPs was obtained.

**Optical characterization**

A UV-vis- near infrared spectrophotometer (Cary 5000, US) with an integrating sphere detector was used to detect the independent contributions of the scattering and absorption to the extinction according to the literature.[36-37] Forward scattering (FS) and backward scattering (BS) signals, which composed the scattering signals (S = FS + BS), were measured separately as follows. The FS signals were collected by placing the sample on a front port with the HMP structure facing the interior of the integrating sphere and the back port of the integrating sphere left open. The BS signals were collected by placing the sample at the back port with the front port open. The transmission (T) signals containing the absorption (A) and BS were collected by placing the sample at the front port but with a standard white board with 100% reflection at the back port. Therefore, A was obtained by A = T – BS.

**Evaluation of the photothermal response**

The temperature increase of the substrate with HMPs was measured by illuminating a 1064 nm laser (pulse length 8 ps, repetition rate 80 MHz) onto the samples with 200 mW power and a 5 mm spot diameter. A thermal camera (Fluke Ti200) was used to monitor the temperature increase of the substrate. The starting temperature is the room temperature. The temperature changes were recorded with an interval of 15 s.

**Cell culture**

NIH-3T3 cells were cultured on both the substrate with HMPs and the gold nano disk control sample. Before seeding the cells, 20 min of UV irradiation was performed in a laminar-flow hood to sterilize the substrate. Subsequently, NIH-3T3 cells were seeded on the devices at a concentration of $0.7 \times 10^4$ cells/cm$^2$ and incubated at 37°C in a 5% $CO_2$ atmosphere for 24 h in DMEM with 1% pen/strep antibiotics and 10% foetal bovine serum (Sigma Aldrich).

**Cell ablation**

Before the cell ablation test, each substrate with cells on top was transferred to a glass bottom petri dish with culture medium. Calcein-AM and propidium iodide (PI) (Sigma Aldrich) were added to perform a live/dead test. A 1064 nm laser (pulse length 8p s, repetition rate 80 MHz), was focused on the substrate with a power of 200 mW and a diameter of 350 μm. After 5 min of laser illumination, images of the live and dead cells were collected from the bottom side of the sample with a 10× objective.

## Results and Discussion

We first designed a meta-particle to maximize the optical absorption at 1064 nm while maintaining low scattering. Absorption efficiency is defined as: $Q_{Abs} = A/(A+S) = A/E$, where $A$ indicates the absorption, $S$ indicates, the scattering and $E$ indicates the total extinction.[18] Since $Q_{Abs}$ is proportional to the heat conversion efficiency (see Supplementary Information S1 for details), the latter variable is widely used to evaluate the heat generating ability of nanoparticles [12, 25, 38]; additionally, increasing the absorption efficiency will improve the heat conversion performance of the nanoparticles.

Toward this aim, we firstly designed the sample working in the NIR II window with the help of numerical simulations, which were implemented using a finite element method (see Methods for more details). As indicated in Figure 1(a), the main concept is as follows: by dividing a gold plasmonic disk into small slices and separating these slices with a dielectric material, one can actually increase the absorption cross section of the system by almost one order of magnitude. Meanwhile, the scattering cross section is reduced by at least a factor of 2.5. Therefore, the absorption efficiency of the HMP can also be increased significantly, as shown in Figure 1(b). This is possible because the intrinsic multi-layered structure of the HMPs can be modified to fully control the scattering and absorption ratio.[33]

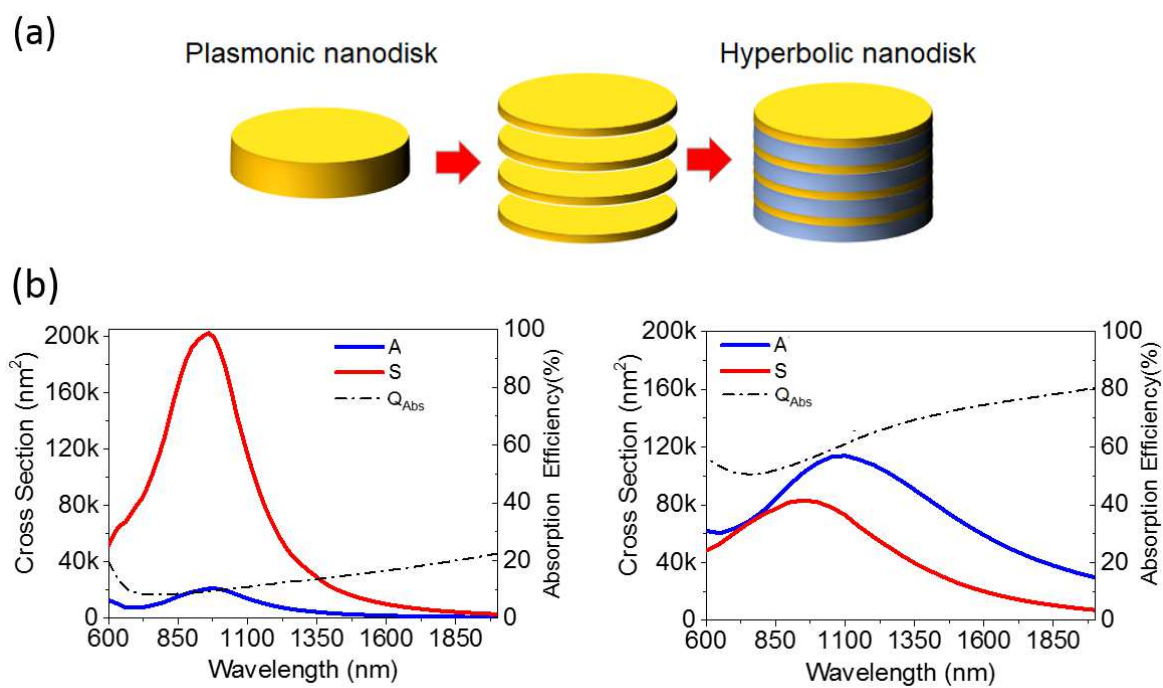

*Figure 1. Concept of the HMP: starting from a pure gold nanodisk (a, left panel) with a strong scattering cross section (b, left panel), a hyperbolic nanodisk with the same amount of gold but nanostructured in a different way (4 bilayers of gold/silica, a, right panel) can increase the absorption cross section (blue line), decrease the scattering cross section (red line) and increase the absorption efficiency (black dashed line) (b, right panel).*

Remarkably, the designed system shows an absorption efficiency that constantly growing the NIR II and III, reaching a value of 80%. Guided by the simulation results, HMP samples composed of four Au(10 nm)/SiO$_2$ (20 nm) bilayers (Figure 2) were fabricated by employing the top down etching method (see Methods and Figure S1). Our method allows for large-area fabrication and an optimal homogeneous distribution of the HMPs on the glass substrate. A Ti (2 nm) adhesion layer was added between the Au and SiO$_2$ to maintain the rod shape. The HMPs were strong enough to maintain their shape after soaking in phosphate-buffered saline for 24 h (Figure S2). As control samples for a performance comparison, we also fabricated gold nanodisks with the same amount of metal (thickness 40 nm) on a glass substrate with diameters and densities similar to the HMPs (Figure S3).

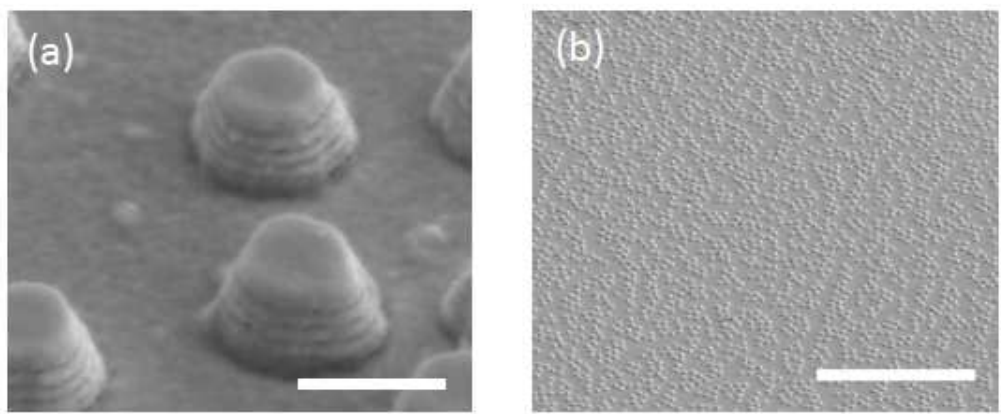

*Figure 2. (a) and (b) SEM images of the four bilayer HMPs on the glass substrate. The scale bars are 300 nm and 10 µm respectively.*

To validate the proposed approach, the absorption and scattering spectra of the HMP particles and gold nanodisks on the glass substrates were measured with a UV-VIS-NIR spectrophotometer with an integrating sphere (see Methods). We chose an average diameter of 230 nm (top disk of the particle) in order to have the absorption peak of both samples be approximately 1064 nm, which is close to the wavelength of the laser for the hyperthermia test.

As shown in Figure 3(a) and (b), The experimental absorption of the HMPs is more than twice that of the pure gold disk in the NIR II window, and the scattering is lower, leading to a higher absorption efficiency. The high absorption of the HMPs is due to the enhanced electric field between the gold layers, which is confined inside the HMP instead of being scattered away (Figure 3(d)). In contrast, such confinements do not appear inside the gold nanodisks, even though the electric field of the gold nanodisks is stronger (Figure 3(b)).

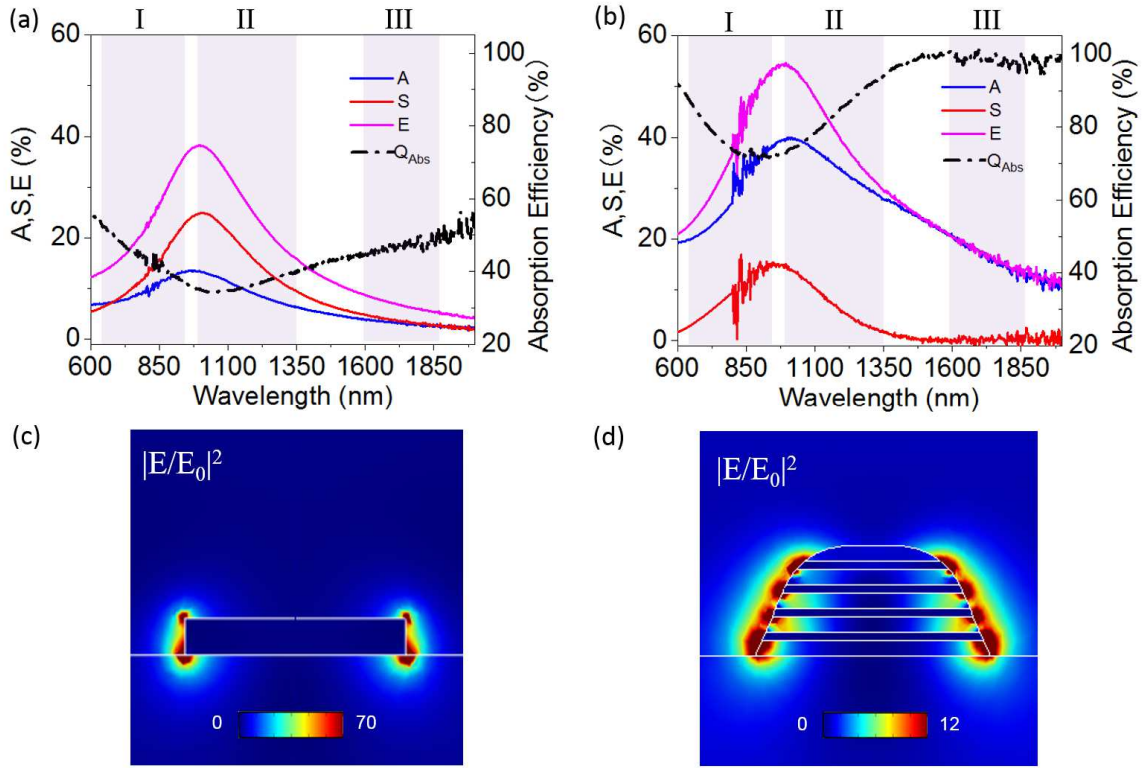

*Figure 3. Experimentally measured Absorption (blue line), Scattering (red line), Extinction (purple line) and Absorption efficiencies (black dashed line) spectrum of (a) gold disks and (b) 4 bi-layer HMPs, respectively. (c) and (d) are simulated near-field distributions of the gold disk and 4 bi-layer HMP, respectively, with an electric intensity enhancement |E/E0|² at a wavelength 1064 nm.*

The HMPs displays pure broad-band absorption in the NIR III window. The scattering of the 4 bi-layer sample is near zero at wavelengths larger than 1500 nm, leading to a high absorption efficiency close to unity at the NIR III window. The absorption of incident optical energy is still lower (12 – 20%) than the LSPR-enhanced absorption (28 – 40%) in the NIR II region. Nevertheless, by changing the thickness and number of the gold/dielectric layer, the HMP LSPR can be tuned to the NIR III region and shows enhanced absorption.[33] Based on the high absorption of the HMPs, we expect a large temperature increase for the HMPs at the NIR II windows. The light-induced temperature increase of nanoparticles is proportional to the adsorption:[39]

$$\Delta T \propto \frac{I_0 A}{nC}$$

where $I_0$ is the incident light intensity, $n$ is the number of nanoparticles in the light spot, and $C$ is the heat capacity of the nanoparticles. To evaluate the thermal response of the system, a temperature increase test was performed in the atmosphere under laser illumination (λ=1064 nm, power=200 mW, spot diameter 5 mm, power density ~1 W/cm$^2$). As shown in Figure 4(a), after 3 min of irradiation, the HMPs reached a temperature increase of 16°C against an increase of only 2°C shown by the gold nanodisk control sample. Hence, the HMP system provides a temperature increase 8 times higher than that of the gold disks systems under the same conditions. Simulated near-field maps of the temperature increase show 5-fold enhancement of the HMP (Figure 4(b)) over the gold nanodisk (Figure 4(c)). The simulations qualitatively agree with the experiments, and its difference from the experiment might come from the far-field measurement setup of temperature (see Methods), as well as the dissipation and convection in the measurements. In Figure 4(d), we plot the magnetic field enhancement as well as the current density. As can be inferred by looking at this figure, a strong current is circulating on the x-y plane on the top of the HMP, thus inducing a strong magnetic field enhancement. These nanoscale optical features lead to a higher temperature increase. Moreover, while the temperature distribution of the gold nanodisk is homogeneous around the disk, the HMP exhibits a high temperature increase on the top side of the structure due to a magnetic field enhancement. Such feature will be useful for hyperthermia applications, as shown below.

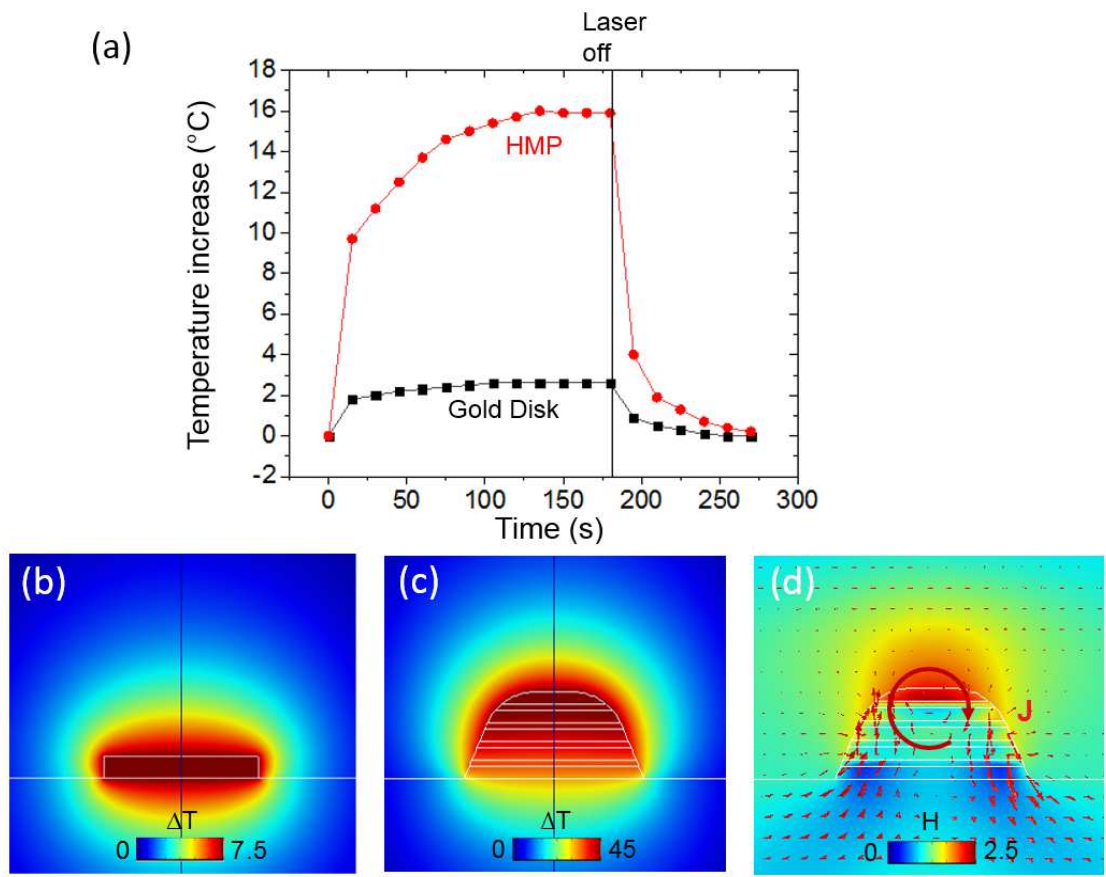

*Figure 4. (a) Experimental temperature increase and decrease with time on the substrate with gold disks and HMPs under 1064 nm laser illumination. At 180 s, the laser was turned off. Calculated steady-state temperature increase of (b) a gold disk on glass and (c) the HMP on the glass substrate upon laser (λ=1064 nm) illumination. (d) Corresponding near-field distribution of the magnetic field enhancement and current density J in HMP, calculated at 1064 nm.*

To validate the *in vitro* hyperthermia performance of the HMPs, we conducted cell ablation tests on substrates with both the HMPs and gold nanodisks by culturing the cells on the top of a substrate patterned with either HMPs or gold nanodisks. This way, a rapid characterization can be performed to avoid the potential problems of cell-nanoparticle interaction, such as endocytosis. Considering that, being a very first characterization of this nanomaterial we still don not know how these nanosystems interacts with cells. For instance, a higher spontaneous cellular uptake with respect to standard gold disks may produce a higher cell mortality, causing misleading interpretations (performance over- or under-estimation). Hence, we decided to use a more conservative approach. After 24 h of cell cultivation, the *NIH-3T3* cells grew well on the HMPs, and most of the cells remained alive; thus, the HMP did not show any obvious cytotoxicity. In the cell ablation test, the incident laser power was 200 mW and was focused to a spot size of 350 µm within the field of the microscope camera monitoring area at a density of 200 W/cm$^2$. After 5 min of laser illumination, the cells on the gold nanodisk remained almost unchanged (Figure 5(a)), whereas those on the HMPs within the laser spot area were mostly dead (Figure 5(b)). With the same amount of gold, the HMPs demonstrated a much better hyperthermia performance than the pure gold nanodisks due to the strong absorption of light in the NIR II window.

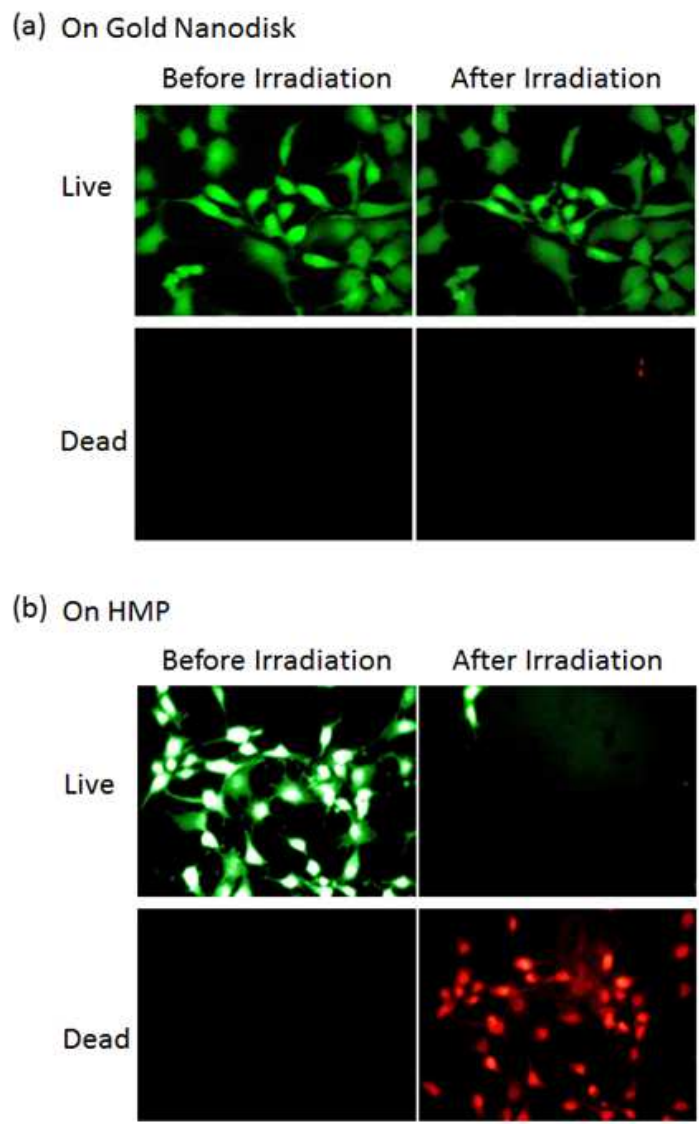

*Figure 5 NIH-3T3 cell ablation test on the substrate. After 5 min of laser irradiation, the cells on (a) the gold nano disks stayed almost unchanged; however, cells on (b) the HMPs were almost dead within the laser spot.*

We remark the importance of introducing this novel class of optical material into bio-related fields and medical applications. In fact, their optical properties can be widely manipulated in order to achieve the desired effect in opposition to conventional nanostructures, which offer less flexibility. Similar concepts can be exploited for antennas or nanostructures that are more sophisticated than a simple gold disk, such as multi-core shell nanoparticles[40], nano-rods, or nanoprisms. Importantly, these composite materials can be further engineered to manipulate not only the optical properties but also the mechanical and chemical ones. For instance, soft polymers can replace the dielectric layers, thus realizing a multi-composite nanoparticle. Drugs or other biomolecules can be loaded into the polymeric material and released upon dissolution. Release can even be triggered on demand upon light stimulation. Additionally, it would be interesting to investigate the biological response to such multi-layered structures and to determine if (and eventually, how) the cells recognize/respond to nano-objects that present periodicity[41][42], e.g., nanomaterials with alternating physico-chemical parameters (for instance, hydrophobic/hydrophilic layers).

## Conclusion

In summary, we showed that multi-layered nanomaterials can provide a straightforward approach to enhancing optical absorption while minimizing the scattering throughout all of the NIR bands. As result, both the absorption efficiency and light-to-heat conversion efficiency are strongly amplified. We designed, fabricated and characterized HMP nanostructures with resonance centred at approximately $\lambda$=1064 nm, which is the excitation wavelength available in our setup. In that spectral range (NIR II), when using HMPs, the light-to-heat conversion efficiency is enhanced by a factor of 8 with respect to its conventional counterpart (gold disk). Importantly, the system can be easily designed to work at even longer wavelengths, thus accessing the NIR III region, which is the most promising spectral range for many medical applications. Within this context, we conducted a preliminary evaluation of how cells (NIH-3T3) grow on multi-layered structures, and we did not find any cytotoxic effect. Successful hyperthermia treatments were achieved in comparison with standard gold disks. The optical response of these systems relies on the hyperbolic behaviour of the dispersion relation. [33,40] The latter can be widely manipulated in order to achieve the desired effect, and similar concepts can be exploited for antennas or nanostructures that are more sophisticated than a simple gold disk. Importantly, this new class of multi-composite nanosystems may enable the exploration of novel mechanical and chemical properties that are not achievable with conventional nanostructures. Among them, novel systems that combine hyperthermia and drug delivery can be foreseen. Hence, we think that this work can pave the way for the exploitation of hyperbolic metamaterials in still unexplored biological and medical contexts, which include photoacoustic imaging, thermal therapy and the controlled release of drugs and biomolecules.

**Corresponding Author**

Francesco De Angelis: francesco.deangelis@iit.it

**Author Contributions**

Y.Z. explored and determined the fabrication protocol, measured the sample optical response, performed the temperature increase test and the cell ablation test and wrote the manuscript. M.I. fabricated the samples, characterized the sample morphology and helped in the temperature increase test and cell ablation test. N.M. designed the hyperbolic meta-particles, performed numerical simulations and simulation results analysis. J.A.H. helped with the sample optical response measurement. F.T. was responsible for the optical setup for the temperature increase test and cell ablation test under laser illumination. L.D. and G.M. performed cell culture and assisted with the cell ablation test. T.I. helped with the sample fabrication and provided suggestions. Y.Z., M.I. and J.H. analysed the experimental data. F.D.A. conceived and supervised the entire work. All the authors contributed to writing the manuscript.

# Reference

1.      Baffou, G.; Quidant, R., Thermo-plasmonics: using metallic nanostructures as nano-sources of heat. *Laser & Photonics Reviews* **2013,** *7* (2), 171-187.

2.      Pelaz, B.; Grazu, V.; Ibarra, A.; Magen, C.; del Pino, P.; de la Fuente, J. M., Tailoring the Synthesis and Heating Ability of Gold Nanoprisms for Bioapplications. *Langmuir* **2012,** *28* (24), 8965-8970.

3.      Su, Y. Y.; Wei, X. P.; Peng, F.; Zhong, Y. L.; Lu, Y. M.; Su, S.; Xu, T. T.; Lee, S. T.; He, Y., Gold Nanoparticles-Decorated Silicon Nanowires as Highly Efficient Near-Infrared Hyperthermia Agents for Cancer Cells Destruction. *Nano Letters* **2012,** *12* (4), 1845-1850.

4.      Ntziachristos, V.; Razansky, D., Molecular Imaging by Means of Multispectral Optoacoustic Tomography (MSOT). *Chem. Rev.* **2010,** *110* (5), 2783-2794.

5.      Jiang, Y.; Lee, H. J.; Lan, L.; Tseng, H.-a.; Yang, C.; Man, H.-Y.; Han, X.; Cheng, J.-X., Optoacoustic brains stimulation at submillimeter spatial precision. *bioRxiv* **2018**, 459933.

6.      Kantner, K.; Rejman, J.; Kraft, K. V. L.; Soliman, M. G.; Zyuzin, M. V.; Escudero, A.; del Pino, P.; Parak, W. J., Laterally and Temporally Controlled Intracellular Staining by Light-Triggered Release of Encapsulated Fluorescent Markers. *Chem-Eur J* **2018,** *24* (9), 2098-2102.

7.      Feng, L. Z.; Cheng, L.; Dong, Z. L.; Tao, D. L.; Barnhart, T. E.; Cai, W. B.; Chen, M. W.; Liu, Z., Theranostic Liposomes with HypoxiaActivated Prodrug to Effectively Destruct Hypoxic Tumors Post-Photodynamic Therapy. *Acs Nano* **2017,** *11* (1), 927-937.

8.      Melancon, M. P.; Zhou, M.; Li, C., Cancer Theranostics with Near-Infrared Light-Activatable Multimodal Nanoparticles. *Accounts Chem. Res.* **2011,** *44* (10), 947-956.

9.      Wang, W. Q.; Wang, L.; Li, Y.; Liu, S.; Xie, Z. G.; Jing, X. B., Nanoscale Polymer Metal-Organic Framework Hybrids for Effective Photothermal Therapy of Colon Cancers. *Adv. Mater.* **2016,** *28* (42), 9320-+.

10.     Espinosa, A.; Di Corato, R.; Kolosnjaj-Tabi, J.; Flaud, P.; Pellegrino, T.; Wilhelm, C., Duality of Iron Oxide Nanoparticles in Cancer Therapy: Amplification of Heating Efficiency by Magnetic Hyperthermia and Photothermal Bimodal Treatment. *Acs Nano* **2016,** *10* (2), 2436-2446.

11.     Wang, S. H.; Riedinger, A.; Li, H. B.; Fu, C. H.; Liu, H. Y.; Li, L. L.; Liu, T. L.; Tan, L. F.; Barthel, M. J.; Pugliese, G.; De Donato, F.; D'Abbusco, M. S.; Meng, X. W.; Manna, L.; Meng, H.; Pellegrino, T., Plasmonic Copper Sulfide Nanocrystals Exhibiting Near-Infrared Photothermal and Photodynamic Therapeutic Effects. *Acs Nano* **2015,** *9* (2), 1788-1800.

12.     Wu, Z. C.; Li, W. P.; Luo, C. H.; Su, C. H.; Yeh, C. S., Rattle-Type Fe3O4@CuS Developed to Conduct Magnetically Guided Photoinduced Hyperthermia at First and Second NIR Biological Windows. *Advanced Functional Materials* **2015,** *25* (41), 6527-6537.

13.     Tsai, M. F.; Chang, S. H. G.; Cheng, F. Y.; Shanmugam, V.; Cheng, Y. S.; Su, C. H.; Yeh, C. S., Au Nanorod Design as Light-Absorber in the First and Second Biological Near-Infrared Windows for in Vivo Photothermal Therapy. *Acs Nano* **2013,** *7* (6), 5330-5342.

14.     Smith, A. M.; Mancini, M. C.; Nie, S. M., BIOIMAGING Second window for in vivo imaging. *Nature Nanotechnology* **2009,** *4* (11), 710-711.

15.     Ding, X. G.; Liow, C. H.; Zhang, M. X.; Huang, R. J.; Li, C. Y.; Shen, H.; Liu, M. Y.; Zou, Y.; Gao, N.; Zhang, Z. J.; Li, Y. G.; Wang, Q. B.; Li, S. Z.; Jiang, J., Surface Plasmon Resonance Enhanced Light Absorption and Photothermal Therapy in the Second Near-Infrared Window. *J. Am. Chem. Soc.* **2014,** *136* (44), 15684-15693.

16.     Bashkatov, A. N.; Genina, E. A.; Kochubey, V. I.; Tuchin, V. V., Optical properties of human skin, subcutaneous and mucous tissues in the wavelength range from 400 to 2000 nm. *J. Phys. D-Appl. Phys.* **2005,** *38* (15), 2543-2555.

17.	Shi L; Sordillo LA; Rodríguez-Contreras A; R, A., Transmission in near-infrared optical windows for deep brain imaging. *Journal of BIOPHOTOMICS* **2016,** *9,* 38-43.

18.	Onal, E. D.; Guven, K., Plasmonic Photothermal Therapy in Third and Fourth Biological Windows. *Journal of Physical Chemistry C* **2017,** *121* (1), 684-690.

19.	Wang, X.; Ma, Y. C.; Sheng, X.; Wang, Y. C.; Xu, H. X., Ultrathin Polypyrrole Nanosheets via Space-Confined Synthesis for Efficient Photothermal Therapy in the Second Near-Infrared Window. *Nano Letters* **2018,** *18* (4), 2217-2225.

20.	Park, J. E.; Kim, M.; Hwang, J. H.; Nam, J. M., Golden Opportunities: Plasmonic Gold Nanostructures for Biomedical Applications based on the Second Near-Infrared Window. *Small Methods* **2017,** *1* (3).

21.	Maestro, L. M.; Haro-Gonzalez, P.; del Rosal, B.; Ramiro, J.; Caamano, A. J.; Carrasco, E.; Juarranz, A.; Sanz-Rodriguez, F.; Sole, J. G.; Jaque, D., Heating efficiency of multi-walled carbon nanotubes in the first and second biological windows. *Nanoscale* **2013,** *5* (17), 7882-7889.

22.	Zhou, J. J.; Jiang, Y. Y.; Hou, S.; Upputuri, P. K.; Wu, D.; Li, J. C.; Wang, P.; Zhen, X.; Pramanik, M.; Pu, K. Y.; Duan, H. W., Compact Plasmonic Blackbody for Cancer Theranosis in the Near-Infrared II Window. *Acs Nano* **2018,** *12* (3), 2643-2651.

23.	Zhang, Q.; Ge, J. P.; Pham, T.; Goebl, J.; Hu, Y. X.; Lu, Z.; Yin, Y. D., Reconstruction of Silver Nanoplates by UV Irradiation: Tailored Optical Properties and Enhanced Stability. *Angewandte Chemie-International Edition* **2009,** *48* (19), 3516-3519.

24.	Cao, Y. Y.; Dou, J. H.; Zhao, N. J.; Zhang, S. M.; Zheng, Y. Q.; Zhang, J. P.; Wang, J. Y.; Pei, J.; Wang, Y. P., Highly Efficient NIR-II Photothermal Conversion Based on an Organic Conjugated Polymer. *Chemistry of Materials* **2017,** *29* (2), 718-725.

25.	Huang, P.; Rong, P. F.; Lin, J.; Li, W. W.; Yan, X. F.; Zhang, M. G.; Nie, L. M.; Niu, G.; Lu, J.; Wang, W.; Chen, X. Y., Triphase Interface Synthesis of Plasmonic Gold Bellflowers as Near-Infrared Light Mediated Acoustic and Thermal Theranostics. *J. Am. Chem. Soc.* **2014,** *136* (23), 8307-8313.

26.	Chen, J. Y.; Glaus, C.; Laforest, R.; Zhang, Q.; Yang, M. X.; Gidding, M.; Welch, M. J.; Xia, Y. N., Gold Nanocages as Photothermal Transducers for Cancer Treatment. *Small* **2010,** *6* (7), 811-817.

27.	Maier, S. A., *Plasmonics: Fundamentals and Applications.* Springer US: 2007.

28.	Jiang, L. Y.; Yin, T. T.; Dong, Z. G.; Hu, H. L.; Liao, M. Y.; Allioux, D.; Tan, S. J.; Goh, X. M.; Li, X. Y.; Yang, J. K. W.; Shen, Z. X., Probing Vertical and Horizontal Plasmonic Resonant States in the Photoluminescence of Gold Nanodisks. *ACS Photonics* **2015,** *2* (8), 1217-1223.

29.	Huang, J. A.; Luo, L. B., Low-Dimensional Plasmonic Photodetectors: Recent Progress and Future Opportunities. *Adv Opt Mater* **2018,** *6* (8).

30.	Olson, J.; Dominguez-Medina, S.; Hoggard, A.; Wang, L. Y.; Chang, W. S.; Link, S., Optical characterization of single plasmonic nanoparticles. *Chem. Soc. Rev.* **2015,** *44* (1), 40-57.

31.	Vijayaraghavan, P.; Liu, C. H.; Vankayala, R.; Chiang, C. S.; Hwang, K. C., Designing Multi-Branched Gold Nanoechinus for NIR Light Activated Dual Modal Photodynamic and Photothermal Therapy in the Second Biological Window. *Adv. Mater.* **2014,** *26* (39), 6689-6695.

32.	Kildishev, A. V.; Boltasseva, A.; Shalaev, V. M., Planar Photonics with Metasurfaces. *Science* **2013,** *339* (6125).

33.	Maccaferri, N.; Zhao, Y.; Isoniemi, T.; Iarossi, M.; Parracino, A.; Strangi, G.; De Angelis, F., Hyperbolic Meta-Antennas Enable Full Control of Scattering and Absorption of Light. *Nano Letters* **2019,** *19* (3), 1851-1859.

34.     Rakic, A. D.; Djurisic, A. B.; Elazar, J. M.; Majewski, M. L., Optical properties of metallic films for vertical-cavity optoelectronic devices. *Appl Optics* **1998,** *37* (22), 5271-5283.

35.     Fredriksson, H.; Alaverdyan, Y.; Dmitriev, A.; Langhammer, C.; Sutherland, D. S.; Zaech, M.; Kasemo, B., Hole-mask colloidal lithography. *Adv. Mater.* **2007,** *19* (23), 4297-+.

36.     Langhammer, C.; Kasemo, B.; Zoric, I., Absorption and scattering of light by Pt, Pd, Ag, and Au nanodisks: Absolute cross sections and branching ratios. *Journal of Chemical Physics* **2007,** *126* (19).

37.     Riley, C. T.; Smalley, J. S. T.; Brodie, J. R. J.; Fainman, Y.; Sirbuly, D. J.; Liu, Z. W., Near-perfect broadband absorption from hyperbolic metamaterial nanoparticles. *P Natl Acad Sci USA* **2017,** *114* (6), 1264-1268.

38.     Du, J. F.; Zheng, X. P.; Yong, Y.; Yu, J.; Dong, X. H.; Zhang, C. Y.; Zhou, R. Y.; Li, B.; Yan, L.; Chen, C. Y.; Gu, Z. J.; Zhao, Y. L., Design of TPGS-functionalized Cu3BiS3 nanocrystals with strong absorption in the second near-infrared window for radiation therapy enhancement. *Nanoscale* **2017,** *9* (24), 8229-8239.

39.     Roper, D. K.; Ahn, W.; Hoepfner, M., Microscale heat transfer transduced by surface plasmon resonant gold nanoparticles. *Journal of Physical Chemistry C* **2007,** *111* (9), 3636-3641.

40.     Wang, P.; Krasavin, A. V.; Viscomi, F. N.; Adawi, A. M.; Bouillard, J. S. G.; Zhang, L.; Roth, D. J.; Tong, L. M.; Zayats, A. V., Metaparticles: Dressing Nano-Objects with a Hyperbolic Coating. *Laser & Photonics Reviews* **2018,** *12* (11).

41.     Jiang, Y. W.; Carvalho-de-Souza, J. L.; Wong, R. C. S.; Luo, Z. Q.; Isheim, D.; Zuo, X. B.; Nicholls, A. W.; Jung, I. W.; Yue, J. P.; Liu, D. J.; Wang, Y. C.; De Andrade, V.; Xiao, X. H.; Navrazhnykh, L.; Weiss, D. E.; Wu, X. Y.; Seidman, D. N.; Bezanilla, F.; Tian, B. Z., Heterogeneous silicon mesostructures for lipid-supported bioelectric interfaces. *Nature Materials* **2016,** *15* (9), 1023-1030.

42.     Rodrigo, D.; Tittl, A.; Ait-Bouziad, N.; John-Herpin, A.; Limaj, O.; Kelly, C.; Yoo, D.; Wittenberg, N. J.; Oh, S. H.; Lashuel, H. A.; Altug, H., Resolving molecule-specific information in dynamic lipid membrane processes with multi-resonant infrared metasurfaces. *Nature Communications* **2018,** *9.*

# Hyperbolic Metamaterial Nanoparticles for Photothermal Therapy in the Second Near-Infrared Window

Yingqi Zhao[1], Marzia Iarossi[1,2], Nicolò Maccaferri[1], Lieselot Deleye[1], Giovanni Melle[1,2],
Jian-An Huang[1], Francesco Tantussi[1], Tommi Isoniemi[1] and Francesco De Angelis[1]*

1 Istituto Italiano di Tecnologia, Via Morego 30, 16163, Genova, Italy
2 Department of Informatics, Bioengineering, Robotics and Systems Engineering (DIBRIS),
Università degli Studi di Genova, Via Balbi 5, 16126, Genova, Italy
*Email: francesco.deangelis@iit.it

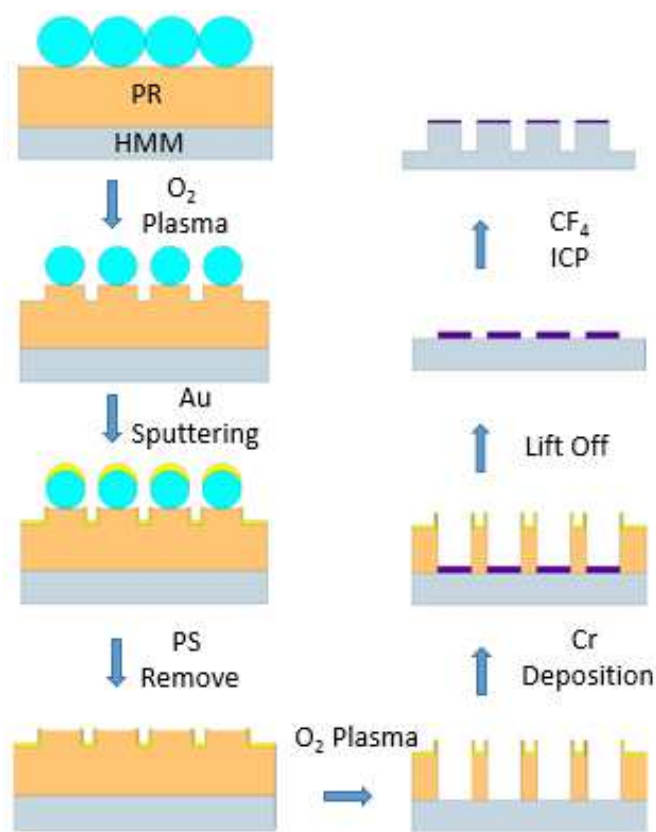

*Figure S1. Fabrication flow chart of HMM Particles*

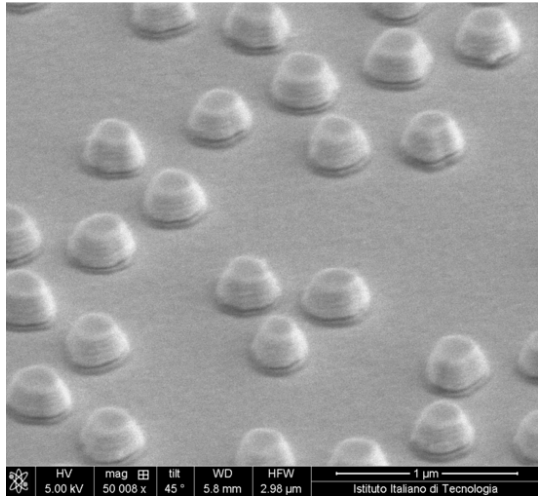

*Figure S2. HMM NPs after soaking in phosphate-buffered saline for 24 h.*

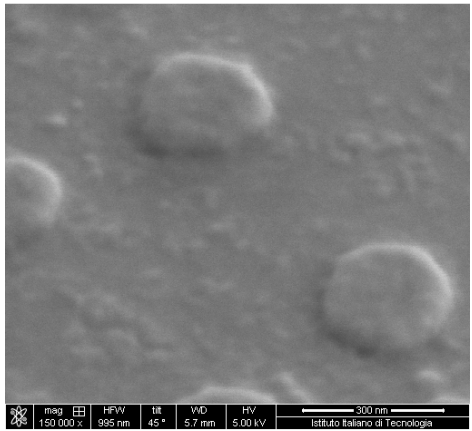

*Figure S3. SEM image of gold disks on glass.*

### S1. Absorption Efficiency and Heat Conversion Efficiency

Absorption efficiency ($Q_{Abs}$) is defined as:[18]

$$Q_{Abs} = \frac{A}{A+S} = \frac{A}{E} \qquad (1)$$

in which A indicates the absorption, S indicates the scattering, and E indicates the extinction.

The heat conversion efficiency is widely used to indicate the nanoparticle performance and can be defined as: [12, 19, 25, 38]

$$\eta = \frac{P}{I_0(1 - 10^{-Abs})} \qquad (2)$$

in which P is the heat generated by the nanoparticles, $I_0$ is the incident light intensity and Abs is the absorbance. The relation between absorbance and transmission is:

$$Abs = -\log_{10}\frac{I}{I_0} = -\log_{10} T \qquad (3)$$

in which I is the transmitted light energy, $I_0$ is the incident light energy and T is transmission. Considering $E = 1 - T = 1 - 10^{-Abs}$, combine equation (2) and (3):

$$\eta = \frac{P}{I_0(1 - 10^{-Abs})} = \frac{P}{I_0 E} \qquad (4)$$

The light-induced temperature rise of nanoparticles is proportional to the adsorption as:[39]

$$\Delta T \propto \frac{I_0 A}{nC} \qquad (5)$$

in which $I_0$ is the incident light intensity, n is the number of nanoparticles in the light spot, and C is the heat capacity of the nanoparticles. Accordingly, the heat generated by the nanoparticle within the laser spot is:

$$P \propto nC\Delta T \qquad (6)$$

Therefore,

$$P \propto I_0 A \qquad (7)$$

Thus, it is obvious that the heat is actually contributed to by the light absorption of the nanoparticles. Combining (4) and (7):

$$\eta = \frac{P}{I_0 E} \propto \frac{A}{E} \qquad (8)$$

With the definition of $Q_{Abs}$, we obtain:

$$\eta \propto Q_{Abs} \qquad (9)$$

## References

1.      Baffou, G.; Quidant, R., Thermo-plasmonics: using metallic nanostructures as nano-sources of heat. *Laser & Photonics Reviews* **2013,** *7* (2), 171-187.

2.      Pelaz, B.; Grazu, V.; Ibarra, A.; Magen, C.; del Pino, P.; de la Fuente, J. M., Tailoring the Synthesis and Heating Ability of Gold Nanoprisms for Bioapplications. *Langmuir* **2012,** *28* (24), 8965-8970.

3.      Su, Y. Y.; Wei, X. P.; Peng, F.; Zhong, Y. L.; Lu, Y. M.; Su, S.; Xu, T. T.; Lee, S. T.; He, Y., Gold Nanoparticles-Decorated Silicon Nanowires as Highly Efficient Near-Infrared Hyperthermia Agents for Cancer Cells Destruction. *Nano Letters* **2012,** *12* (4), 1845-1850.

4.      Ntziachristos, V.; Razansky, D., Molecular Imaging by Means of Multispectral Optoacoustic Tomography (MSOT). *Chem. Rev.* **2010,** *110* (5), 2783-2794.

5.      Jiang, Y.; Lee, H. J.; Lan, L.; Tseng, H.-a.; Yang, C.; Man, H.-Y.; Han, X.; Cheng, J.-X., Optoacoustic brains stimulation at submillimeter spatial precision. *bioRxiv* **2018,** 459933.

6.      Kantner, K.; Rejman, J.; Kraft, K. V. L.; Soliman, M. G.; Zyuzin, M. V.; Escudero, A.; del Pino, P.; Parak, W. J., Laterally and Temporally Controlled Intracellular Staining by Light-Triggered Release of Encapsulated Fluorescent Markers. *Chem-Eur J* **2018,** *24* (9), 2098-2102.

7.      Feng, L. Z.; Cheng, L.; Dong, Z. L.; Tao, D. L.; Barnhart, T. E.; Cai, W. B.; Chen, M. W.; Liu, Z., Theranostic Liposomes with HypoxiaActivated Prodrug to Effectively Destruct Hypoxic Tumors Post-Photodynamic Therapy. *Acs Nano* **2017,** *11* (1), 927-937.

8.      Melancon, M. P.; Zhou, M.; Li, C., Cancer Theranostics with Near-Infrared Light-Activatable Multimodal Nanoparticles. *Accounts Chem. Res.* **2011,** *44* (10), 947-956.

9.      Wang, W. Q.; Wang, L.; Li, Y.; Liu, S.; Xie, Z. G.; Jing, X. B., Nanoscale Polymer Metal-Organic Framework Hybrids for Effective Photothermal Therapy of Colon Cancers. *Adv. Mater.* **2016,** *28* (42), 9320-+.

10.     Espinosa, A.; Di Corato, R.; Kolosnjaj-Tabi, J.; Flaud, P.; Pellegrino, T.; Wilhelm, C., Duality of Iron Oxide Nanoparticles in Cancer Therapy: Amplification of Heating Efficiency by Magnetic Hyperthermia and Photothermal Bimodal Treatment. *Acs Nano* **2016,** *10* (2), 2436-2446.

11.     Wang, S. H.; Riedinger, A.; Li, H. B.; Fu, C. H.; Liu, H. Y.; Li, L. L.; Liu, T. L.; Tan, L. F.; Barthel, M. J.; Pugliese, G.; De Donato, F.; D'Abbusco, M. S.; Meng, X. W.; Manna, L.; Meng, H.; Pellegrino, T., Plasmonic Copper Sulfide Nanocrystals Exhibiting Near-Infrared Photothermal and Photodynamic Therapeutic Effects. *Acs Nano* **2015,** *9* (2), 1788-1800.

12.     Wu, Z. C.; Li, W. P.; Luo, C. H.; Su, C. H.; Yeh, C. S., Rattle-Type Fe3O4@CuS Developed to Conduct Magnetically Guided Photoinduced Hyperthermia at First and Second NIR Biological Windows. *Advanced Functional Materials* **2015,** *25* (41), 6527-6537.

13. Tsai, M. F.; Chang, S. H. G.; Cheng, F. Y.; Shanmugam, V.; Cheng, Y. S.; Su, C. H.; Yeh, C. S., Au Nanorod Design as Light-Absorber in the First and Second Biological Near-Infrared Windows for in Vivo Photothermal Therapy. *Acs Nano* **2013,** *7* (6), 5330-5342.

14. Smith, A. M.; Mancini, M. C.; Nie, S. M., BIOIMAGING Second window for in vivo imaging. *Nature Nanotechnology* **2009,** *4* (11), 710-711.

15. Ding, X. G.; Liow, C. H.; Zhang, M. X.; Huang, R. J.; Li, C. Y.; Shen, H.; Liu, M. Y.; Zou, Y.; Gao, N.; Zhang, Z. J.; Li, Y. G.; Wang, Q. B.; Li, S. Z.; Jiang, J., Surface Plasmon Resonance Enhanced Light Absorption and Photothermal Therapy in the Second Near-Infrared Window. *J. Am. Chem. Soc.* **2014,** *136* (44), 15684-15693.

16. Bashkatov, A. N.; Genina, E. A.; Kochubey, V. I.; Tuchin, V. V., Optical properties of human skin, subcutaneous and mucous tissues in the wavelength range from 400 to 2000 nm. *J. Phys. D-Appl. Phys.* **2005,** *38* (15), 2543-2555.

17. Shi L; Sordillo LA; Rodríguez-Contreras A; R, A., Transmission in near-infrared optical windows for deep brain imaging. *Journal of BIOPHOTONICS* **2016,** *9*, 38-43.

18. Onal, E. D.; Guven, K., Plasmonic Photothermal Therapy in Third and Fourth Biological Windows. *Journal of Physical Chemistry C* **2017,** *121* (1), 684-690.

19. Wang, X.; Ma, Y. C.; Sheng, X.; Wang, Y. C.; Xu, H. X., Ultrathin Polypyrrole Nanosheets via Space-Confined Synthesis for Efficient Photothermal Therapy in the Second Near-Infrared Window. *Nano Letters* **2018,** *18* (4), 2217-2225.

20. Park, J. E.; Kim, M.; Hwang, J. H.; Nam, J. M., Golden Opportunities: Plasmonic Gold Nanostructures for Biomedical Applications based on the Second Near-Infrared Window. *Small Methods* **2017,** *1* (3).

21. Maestro, L. M.; Haro-Gonzalez, P.; del Rosal, B.; Ramiro, J.; Caamano, A. J.; Carrasco, E.; Juarranz, A.; Sanz-Rodriguez, F.; Sole, J. G.; Jaque, D., Heating efficiency of multi-walled carbon nanotubes in the first and second biological windows. *Nanoscale* **2013,** *5* (17), 7882-7889.

22. Zhou, J. J.; Jiang, Y. Y.; Hou, S.; Upputuri, P. K.; Wu, D.; Li, J. C.; Wang, P.; Zhen, X.; Pramanik, M.; Pu, K. Y.; Duan, H. W., Compact Plasmonic Blackbody for Cancer Theranosis in the Near-Infrared II Window. *Acs Nano* **2018,** *12* (3), 2643-2651.

23. Zhang, Q.; Ge, J. P.; Pham, T.; Goebl, J.; Hu, Y. X.; Lu, Z.; Yin, Y. D., Reconstruction of Silver Nanoplates by UV Irradiation: Tailored Optical Properties and Enhanced Stability. *Angewandte Chemie-International Edition* **2009,** *48* (19), 3516-3519.

24. Cao, Y. Y.; Dou, J. H.; Zhao, N. J.; Zhang, S. M.; Zheng, Y. Q.; Zhang, J. P.; Wang, J. Y.; Pei, J.; Wang, Y. P., Highly Efficient NIR-II Photothermal Conversion Based on an Organic Conjugated Polymer. *Chemistry of Materials* **2017,** *29* (2), 718-725.

25. Huang, P.; Rong, P. F.; Lin, J.; Li, W. W.; Yan, X. F.; Zhang, M. G.; Nie, L. M.; Niu, G.; Lu, J.; Wang, W.; Chen, X. Y., Triphase Interface Synthesis of Plasmonic Gold Bellflowers as Near-Infrared Light Mediated Acoustic and Thermal Theranostics. *J. Am. Chem. Soc.* **2014,** *136* (23), 8307-8313.

26. Chen, J. Y.; Glaus, C.; Laforest, R.; Zhang, Q.; Yang, M. X.; Gidding, M.; Welch, M. J.; Xia, Y. N., Gold Nanocages as Photothermal Transducers for Cancer Treatment. *Small* **2010,** *6* (7), 811-817.

27. Maier, S. A., *Plasmonics: Fundamentals and Applications*. Springer US: 2007.

28. Jiang, L. Y.; Yin, T. T.; Dong, Z. G.; Hu, H. L.; Liao, M. Y.; Allioux, D.; Tan, S. J.; Goh, X. M.; Li, X. Y.; Yang, J. K. W.; Shen, Z. X., Probing Vertical and Horizontal Plasmonic Resonant States in the Photoluminescence of Gold Nanodisks. *ACS Photonics* **2015,** *2* (8), 1217-1223.

29. Huang, J. A.; Luo, L. B., Low-Dimensional Plasmonic Photodetectors: Recent Progress and Future Opportunities. *Adv Opt Mater* **2018,** *6* (8).

30.     Olson, J.; Dominguez-Medina, S.; Hoggard, A.; Wang, L. Y.; Chang, W. S.; Link, S., Optical characterization of single plasmonic nanoparticles. *Chem. Soc. Rev.* **2015,** *44* (1), 40-57.

31.     Vijayaraghavan, P.; Liu, C. H.; Vankayala, R.; Chiang, C. S.; Hwang, K. C., Designing Multi-Branched Gold Nanoechinus for NIR Light Activated Dual Modal Photodynamic and Photothermal Therapy in the Second Biological Window. *Adv. Mater.* **2014,** *26* (39), 6689-6695.

32.     Kildishev, A. V.; Boltasseva, A.; Shalaev, V. M., Planar Photonics with Metasurfaces. *Science* **2013,** *339* (6125).

33.     Maccaferri, N.; Zhao, Y.; Isoniemi, T.; Iarossi, M.; Parracino, A.; Strangi, G.; De Angelis, F., Hyperbolic Meta-Antennas Enable Full Control of Scattering and Absorption of Light. *Nano Letters* **2019,** *19* (3), 1851-1859.

34.     Rakic, A. D.; Djurisic, A. B.; Elazar, J. M.; Majewski, M. L., Optical properties of metallic films for vertical-cavity optoelectronic devices. *Appl Optics* **1998,** *37* (22), 5271-5283.

35.     Fredriksson, H.; Alaverdyan, Y.; Dmitriev, A.; Langhammer, C.; Sutherland, D. S.; Zaech, M.; Kasemo, B., Hole-mask colloidal lithography. *Adv. Mater.* **2007,** *19* (23), 4297-+.

36.     Langhammer, C.; Kasemo, B.; Zoric, I., Absorption and scattering of light by Pt, Pd, Ag, and Au nanodisks: Absolute cross sections and branching ratios. *Journal of Chemical Physics* **2007,** *126* (19).

37.     Riley, C. T.; Smalley, J. S. T.; Brodie, J. R. J.; Fainman, Y.; Sirbuly, D. J.; Liu, Z. W., Near-perfect broadband absorption from hyperbolic metamaterial nanoparticles. *P Natl Acad Sci USA* **2017,** *114* (6), 1264-1268.

38.     Du, J. F.; Zheng, X. P.; Yong, Y.; Yu, J.; Dong, X. H.; Zhang, C. Y.; Zhou, R. Y.; Li, B.; Yan, L.; Chen, C. Y.; Gu, Z. J.; Zhao, Y. L., Design of TPGS-functionalized Cu3BiS3 nanocrystals with strong absorption in the second near-infrared window for radiation therapy enhancement. *Nanoscale* **2017,** *9* (24), 8229-8239.

39.     Roper, D. K.; Ahn, W.; Hoepfner, M., Microscale heat transfer transduced by surface plasmon resonant gold nanoparticles. *Journal of Physical Chemistry C* **2007,** *111* (9), 3636-3641.

40.     Wang, P.; Krasavin, A. V.; Viscomi, F. N.; Adawi, A. M.; Bouillard, J. S. G.; Zhang, L.; Roth, D. J.; Tong, L. M.; Zayats, A. V., Metaparticles: Dressing Nano-Objects with a Hyperbolic Coating. *Laser & Photonics Reviews* **2018,** *12* (11).

41.     Jiang, Y. W.; Carvalho-de-Souza, J. L.; Wong, R. C. S.; Luo, Z. Q.; Isheim, D.; Zuo, X. B.; Nicholls, A. W.; Jung, I. W.; Yue, J. P.; Liu, D. J.; Wang, Y. C.; De Andrade, V.; Xiao, X. H.; Navrazhnykh, L.; Weiss, D. E.; Wu, X. Y.; Seidman, D. N.; Bezanilla, F.; Tian, B. Z., Heterogeneous silicon rnesostructures for lipid-supported bioelectric interfaces. *Nature Materials* **2016,** *15* (9), 1023-1030.

42.     Rodrigo, D.; Tittl, A.; Ait-Bouziad, N.; John-Herpin, A.; Limaj, O.; Kelly, C.; Yoo, D.; Wittenberg, N. J.; Oh, S. H.; Lashuel, H. A.; Altug, H., Resolving molecule-specific information in dynamic lipid membrane processes with multi-resonant infrared metasurfaces. *Nature Communications* **2018,** *9.*